\documentclass[
aps,
12pt,
final,
notitlepage,
oneside,
onecolumn,
nobibnotes,
nofootinbib,
superscriptaddress,
noshowpacs,
centertags]
{revtex4}
\usepackage{graphicx}

\begin{document}

\title{Do low surface brightness galaxies have dense disks?}

\author{\ {A. S.}~\surname{Saburova}}
\email{saburovaann@gmail.com}
\affiliation{
Sternberg Astronomical Institute, Moscow State University}

\begin{abstract}
The disk masses of four low surface brightness galaxies (LSB) were estimated using marginal gravitational stability criterion and the stellar velocity dispersion data which were taken from Pizzella et al., 2008 \cite{Pizzella}. The constructed mass models appear to be close to the models of maximal disk. The results show that the disks of LSB galaxies may be significantly more massive than it is usually accepted from their brightnesses. In this case their surface densities and masses appear to be rather typical for normal spirals. Otherwise, unlike the disks of many spiral galaxies, the LSB disks are dynamically overheated.

\end{abstract}
\maketitle

\section{Introduction}
LSB galaxies represent the wide class of disk galaxies, central surface brightnesses of which are many times lower than those of normal spiral and S0 galaxies. Among LSB galaxies there are both dwarf and giant systems, both objects with solar metallicity and systems with significant deficiency of metals. The gas mass over the luminosity $M_{HI}/L_B$ ratio of LSB galaxies lies in the wide interval: 0,1-50 solar units, $(B-V)$ color indices of these objects also span a wide range of values (O'Neil et al., 2000 \cite{oneil}). All these properties indicate that LSB galaxies are not homogeneous by their origin and the character of evolution.

In general, disks of LSB galaxies resemble disks of high surface brightness galaxies (HSB) with a scaled down density of stellar population and, usually, with lower surface density of gas. They have a similar character of surface brightness radial distribution (brightness radial profiles are close to exponential, and can also have a sharp cut-off, as for many HSB-galaxies). Spiral arms and bars are also observed in some of LSB galaxies (though they occur not so frequently as in HSB ones).  The amount of young stars with high UV-luminosity per unit of $HI$ mass in these systems is by an order of magnitude lower than in normal galaxies (Boissier et al., 2008 \cite{Boissier}).

The most essential property of LSB-galaxies is a high total mass-to-light ratio $M/L$ (within several radial scalelenghts), indicating the great amount of dark matter or low massive stars. However a high total $M/L$ doesn't provide a clue to the location of dark mass, whether it is located in halo, in disk, or in both components.
For clarifying the nature of LSB-galaxies the key role plays the problem of their disk mass. If normal stellar population and gas make up the bulk of a disk mass, the LSB disks have a very low surface density, and the total mass of these galaxies inside of optical radius must be several times higher than that of the baryon matter. The conclusion about the dominating dark matter contribution within optical radii of these galaxies is based on these arguments (see e.g. Bothun et al., 1997 \cite{Bothun} and de Blok, McGaugh, 1997 \cite{Blok97},  de Blok, Bosma, 2002 \cite{Blok02}, Kuzio de Naray et al, 2008 \cite{Kuzio08}). This assumption allows one to derive the dark halo density distribution directly from the rotation curve of LSB-galaxies. A comparison of the obtained density profile with that resulted from numerical simulation of the formation of galaxies may be used to test different scenarios of formation of galaxies in dark halo gravitational field (see e.g. Kuzio de Naray et al, 2008 \cite{Kuzio08}).

The low thickness $z_*$ of edge-on LSB disks, with respect to the radial scalelenght $h$ is also an important argument in favor of dominating of dark halo mass contribution within several $h$ and a marginal stability of their disks (Zasov, Bizyaev, 2003 \cite{Zasov2}, Bizyaev, Mitronova, 2009 \cite{Bizyaev}).

Nevertheless, the question whether dark halo mass dominates in LSB-galaxies and how often it occurs remains open. For LSB-galaxies with measured rotation curves the dominating contribution of dark halo results primarily from the decomposition of the rotation curves, using the surface brightness and color indices estimates, which provide the information about the mass of stellar population. In this case it is usually assumed that the relation between color and $M/L$ ratio is similar to that of HSB-galaxies, what is generally not so evident: disks may be significantly denser, and it will not contradict to the observed kinematics.

The upper limit of a disk density may be found, using the so called maximal disk model, under the condition that a dark halo of a galaxy doesn't posses an inner hollow. $M/L_R$ ratios of stellar population, obtained by this model for LSB-galaxies, appear to span a wide range of values Ц from 1-2 to 10-16 solar units (de Blok, McGaugh, 1997 \cite{Blok97} , de Blok, Bosma, 2002 \cite{Blok02}, Kuzio de Naray et al, 2008 \cite{Kuzio08}), while for normal stellar population it lies in the interval 0,5-3 solar units. Unlike LSB-galaxies, for giant HSB-galaxies with high luminosities the maximal disk model gives the values of $M/L$, differing from those referred to the stellar population synthesis model usually no more than by $0.1 dex$, what is within the uncertainties of photometrical method (see Kassin, de Jong, Weiner, 2006 \cite{Kassin}). Hence it appears that either a mass of LSB disks is significantly lower in comparison with that, expected for the maximal disk model being negligible with respect to the dark halo mass, or the disks are really as massive as in the HSB-galaxies and contain a large amount of invisible matter: low massive stars or other kinds of baryonic (e.g. cold molecular hydrogen) or non-baryonic dark mass, (see de Blok и McGaugh 1997 \cite{Blok97}).

 However, the estimate of maximal disk mass is not reliable, because it strongly depends on the accuracy of determination of the rising part of rotation curve. A perfect examples of such case can be Malin-1 and NGC7589, where two different methods of the $HI$ kinematic data reduction have led to the dramatically different shapes of the inner parts of rotation curves (F.Lelli, R.Sancisi, F.Fraternali, 2010 \cite{Lelli}). The disk component of rotation curve of LSB-galaxies usually has a peak at such distance $R$ from the center, where the observed rotation curve doesn't reach its flat part yet. That's why even a small discrepancy between the radial scalelenghts of surface brightness and density leads to significant errors of estimation of the disk contribution to the rotation curve. It is worth noting, that the obtained value of a disk mass depends also on the assumption of the constancy of stellar population mass-to-light ratio.

The alternative method of the disk mass estimation is based on the analysis of the inner motions or the condition of spiral structure formation in disks. This approach leads to the conclusion that the disks may be more massive than it follows from photometrical data. Masset, Bureau, 2003 \cite{Masset} have considered the numerical dynamical model of a galaxy, reproducing the formation of two-armed spiral structure in the extensive gaseous disk of the dwarf galaxy NGC 2915. Authors came to the conclusion that its disk contained dark matter, which distribution was proportional to that of $HI$ mass, and was by an order of magnitude more massive than the latter one. Fuchs, 2003 \cite{Fuchs03} on the basis of the density wave linear theory has obtained for the disks of several LSB-galaxies with two-armed spiral structure the values of $M/L_R$, lying in the interval 4-16, which were significantly higher than it was predicted by stellar population models, but surprisingly close to the estimates of maximal disk model. The existence of periodical large-scale perturbations in the disk of edge-on LSB-galaxy IC2233, also indicates indirectly that the disk is self-graviting (Matthews, Uson 2008 \cite{Matthews}).

In the work of Lee et al., 2004 \cite{Lee} a high mass-to-light ratios of the disks of LSB-galaxies are explained in terms of abnormally steep initial mass function with a power law exponent of $\alpha \approx 3,85$ (compared with the standard Salpeter IMF $\alpha \approx 2,35$). A lower content of evolved stars in this case explains the lower metallicity, which, parallel with the lower amount of dust, leads to the lower color index of stellar population. Note that the problem with the interpretation of the observed chemical composition of LSB-galaxies with high values of $\alpha$ still remains (Mattsson et al, 2008 \cite{Mattsson}).
	
The independent estimate of the upper limit of disk mass can be obtained on the basis of measured stellar velocity dispersion. This method, firstly applied by Bottema, 1993, 1997 \cite{Bottema93}, \cite{Bottema97} to HSB-galaxies, relies on the assumption of the stability of a disk against radial density perturbations. The denser is the disk, the higher should be the velocity dispersion of stars (or gas, if it makes up the bulk of disk mass) for the disk to be stable. The evaluations of disk mass and mass-to-light ratio for normal spiral galaxies (besides some early type spirals), obtained in such a way, are in good agreement with those referred to the stellar population synthesis models. This enables to conclude, that the disk density of galaxies is usually close to that, expected for the marginal gravitational stability (Zasov et al., 2004, 2010 \cite{Zasov}, \cite{Zasov10})

\section{The used method}
The constraints on the disk surface density distribution may be found on the basis of radial velocity dispersion, using the condition of gravitational stability of stellar disk. The critical value of radial velocity dispersion which makes a thin, collisionless, isothermal disk stable against the density perturbations at all wavelengths, is usually written as:
 \begin{equation}
\label{formula2}c_{cr}=Q_c\cdot c_{T}
\end{equation}
where $Q_c$ is a Toomre stability parameter, $c_{T}\approx \frac{3,4 \cdot G \cdot \sigma _*} {\kappa}$ is Toomre critical radial velocity dispersion,
 $\sigma _*$ is a disk surface density, $\kappa$ is the epicyclical frequency:
\begin{equation}
\label{formula3} \kappa =2\Omega\cdot\sqrt {1+(R/2\Omega)(d\Omega/dR)}
         \end{equation}
$\Omega$ is the angular velocity. Radial velocity dispersion can be found from the line-of-sight velocity dispersion measured along the slit passing through the center of a galaxy using a simple expression:
\begin{equation}
\label{formula4} c_{obs}(r) = (c_z \cdot cos^2 (i)+ c_{\phi}\cdot sin^2  (i)\cdot cos^2(\alpha)+ c_{r}sin^2(i)\cdot sin^2(\alpha))^{0,5}
\end{equation}
and the additional conditions:
$ c_{r} = 2\Omega \cdot  c_{\phi} /\kappa $ and $ c_z =k\cdot c_{r} $, where $c_{obs} (r)$ is the line-of-sight velocity dispersion, $c_z$, $c_{\phi}$, $c_{r}$ are its vertical, azimuthal and radial components, $i$ is the inclination of the disk of galaxy, $\alpha$ is the de-projected angle between the slit and the major axis. The coefficient $k$ was taken to be equal to 0,6.
In the model of a thin disk and axissymmetric perturbations, the Toomre parameter $ Q_c =1$. But in general, the value of $Q_c(R)$ depends on geometrical and kinematical parameters of a disk and on the initial conditions of its dynamical evolution. In the current work the Toomre parameter was taken to vary with radius as it resulted from the dynamical modeling of 3D marginally stable disks (Khoperskov et al., 2003 \cite{Khoperskov}):
\begin{equation}
\label{formula5} Q_c(r/h)  = A_0+A_1\cdot  (r/h)+A_2\cdot  (r/h)^2
  \end{equation}
where $A_0=1,25$,  $A_1=-0,19$,  $A_2=0,134$, and $r/h$ - is the distance from the center of galaxies in the units of a disk scalelength.

\section{Mass-to-light ratios}
The gravitational stability criterion (\ref{formula2}) can be used to calculate the upper limit of the disk local density at a given galactocentric distance $R$, and to determine the maximal contribution of the locally stable disk to the rotation curve. It was done for four LSB-galaxies, for which the distribution of the velocity dispersion and rotation curves of ionized gas and stars, parallel with the photometrical profiles in R-band, were obtained by Pizzella et al., 2008 \cite{Pizzella}
Some parameters of the selected galaxies are listed in Table \ref{table1}:\\
(1) Ц- Name\\
(2) Ц- Distance D in Mpc \\
(3) -- Inclination\\
(4) Ц- Morphological type\\
(5) Ц- Absolute stellar magnitude in B-band \\
(6) Ц- Mass of $HI$, normalized by luminosity $L_B$, (parameter $hic$ in Hyperleda \cite{leda})\\
(7)-- Notes  \\
\begin{table}[h!]
\small \caption{The main properties. \label{table1}}
  \begin{center}
    \begin{tabular}{|c|c|c|c|c|c|c|c|c|}
    \hline
Name& $D$, Mpc & $i$ & Type & $M_B$ &$hic$& Notes\\
 (1) & (2) & (3) & (4) & (5)&(6)&(7) \\
 \hline
ESO-LV 1860550&60,1&63&Sab&-19,9&1,25&-\\
\hline
ESO-LV 2060140&60,5&39&SABc&-19,65&-&in group\\
\hline
ESO-LV 2340130&60,9&69&Sbc&-20&1,11&in group, in cluster \\
\hline
ESO-LV 4000370&37,5&50&SBc&-19,1&	1.39&in cluster\\
\hline
 \end{tabular}
  \end{center}
\end{table}

The criterion (\ref{formula2}) doesn't take into account a contribution of cold gaseous component to the condition of stability of a disk, which leads to the increasing of parameter $Q_c$. However, the surface density of gas, estimated for three of four considered galaxies on the basis of Hyperleda \cite{leda} data (for ESO-LV2060140 such data are not available) is significantly lower than that of stellar disk. Taking into account the gas contribution for such ratio of stellar and gas surface densities following the models of Rafikov, 2001 \cite{Rafikov} and Morozov, Khoperskov, 2005 \cite{Morozov} leads to the conclusion that the overestimation of disk surface density for marginally stable disk does not exceed 25\% of its value. Hence the influence of gaseous component can be neglected.

The ionized gas rotation curves of considered LSB galaxies, obtained by Pizzella et al, 2008 \cite{Pizzella}, are highly asymmetric and irregular in comparison with the stellar ones, in this connection these authors recommend to use the stellar rotation curves in dynamical modeling. But stellar rotation curves are not as extensive as gas ones, so both stellar and gas rotation curves were used in the current work to construct the dynamical models. Stellar rotation curves were corrected for asymmetric drift according to Binney, Tremaine, 1987 \cite{binney}:
\begin{equation}
\label{formula1} v_c^2 = v_r^2 + c_r^2\cdot  (0,5\cdot  d(lnv_r)/d(lnr) - 0,5 +r/h -  d(ln c_r^2)/d(lnr)),
\end{equation}
where $v_c$ and $v_r$ are circular and observed rotation velocities correspondingly, $c_r$ is the radial velocity dispersion, $h$ is radial scalelenght of the disk.

Surface density radial profiles of marginally stable disks, obtained by applying the local criterion (\ref{formula2}), were used to estimate $(M/L_R)_{disk}$ for different galactocentric distances (see Fig.1). The filled and open symbols in Fig.1 correspond to the gas and stellar circular velocities. The dashed lines denote $(M/L_R)_{disk}$ ratios, predicted by stellar population models (Bell, de Jong, 2001 \cite{bdj}) for the metallicity $Z=0,008$, the solid lines correspond to the mean values of $(M/L_R)_{disk}$, which were accepted for the decomposition of rotation curves.
\begin{figure} [h!]
\includegraphics[width=8cm,keepaspectratio]{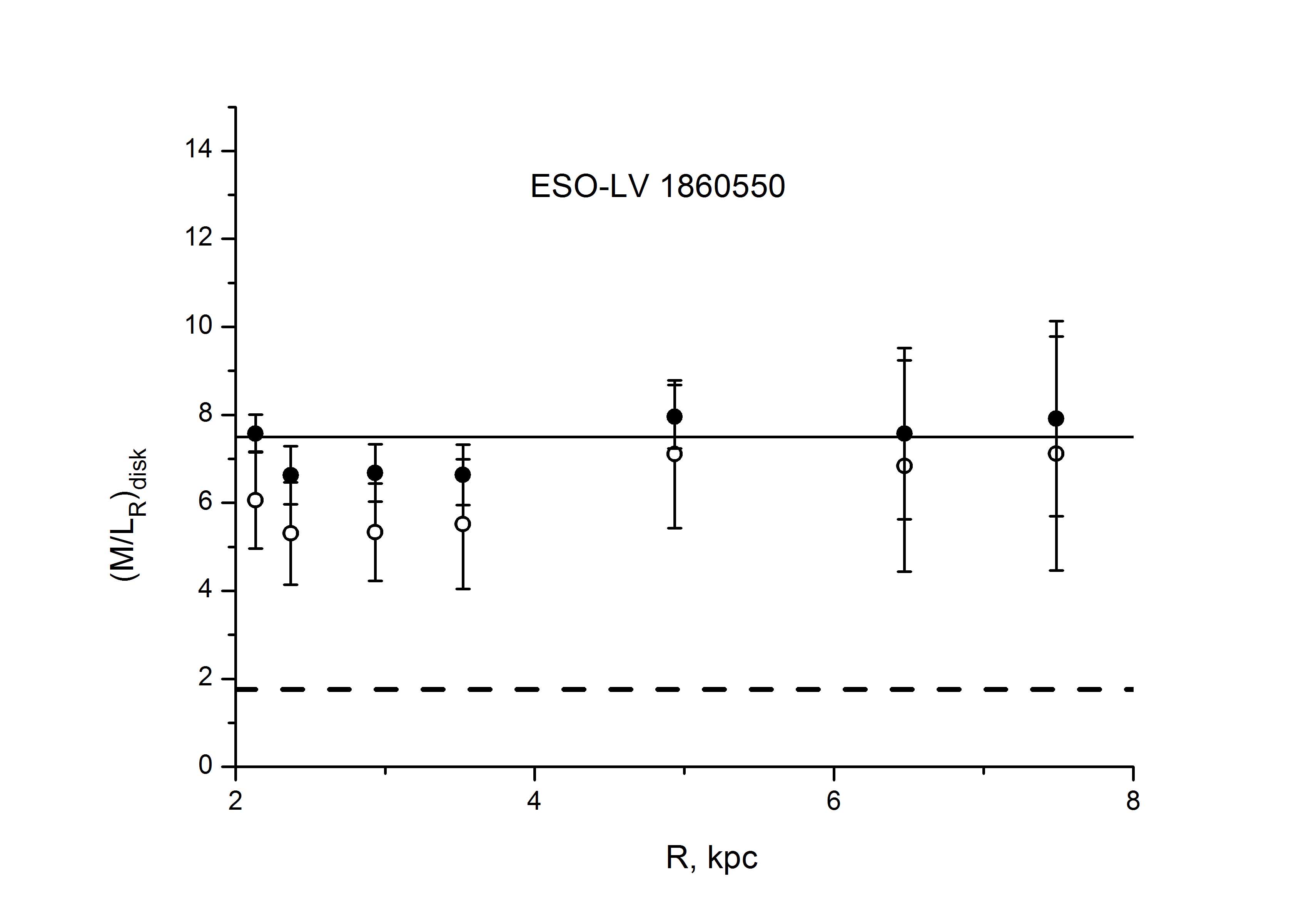}
\includegraphics[width=8cm,keepaspectratio]{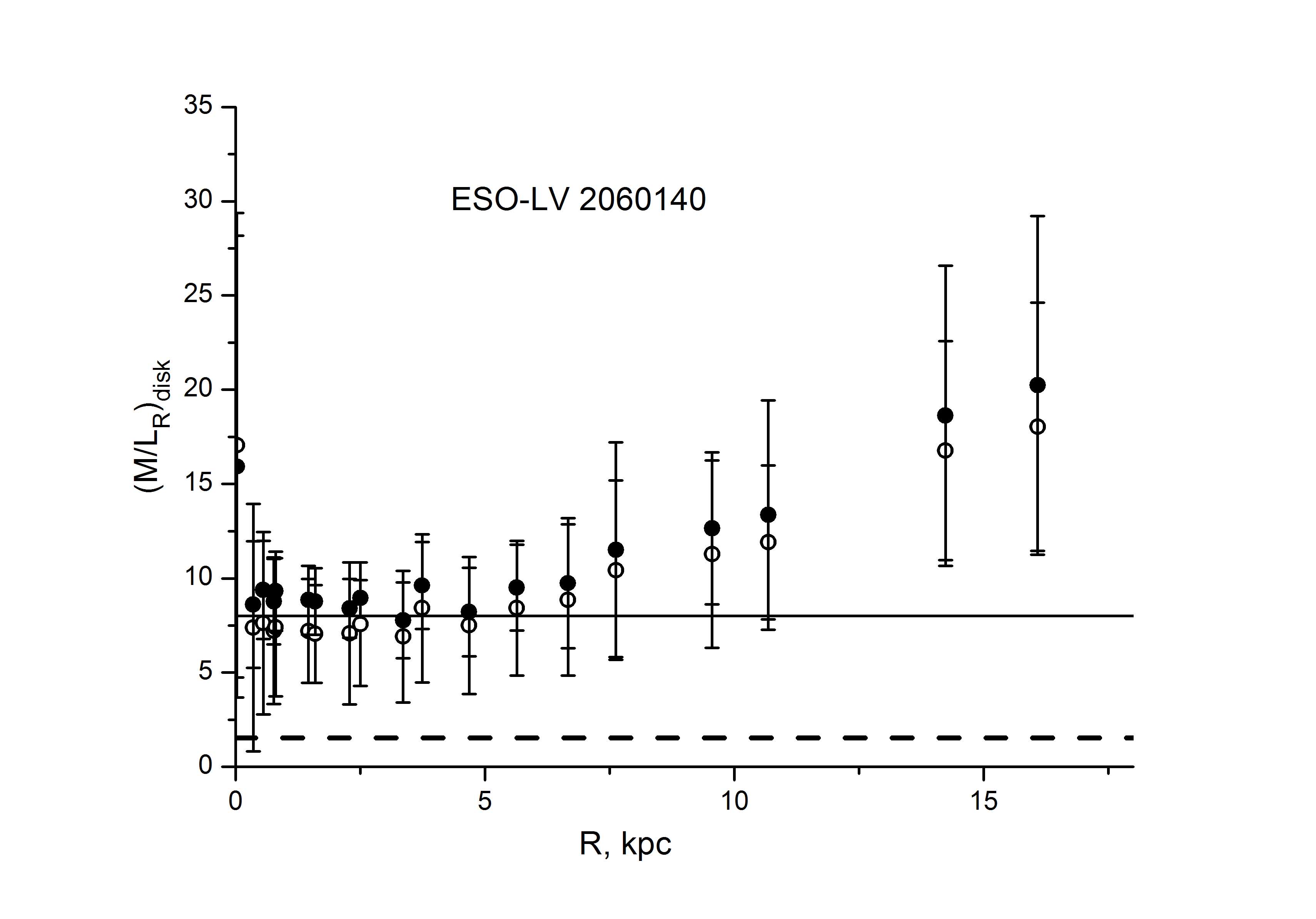}
\includegraphics[width=8cm,keepaspectratio]{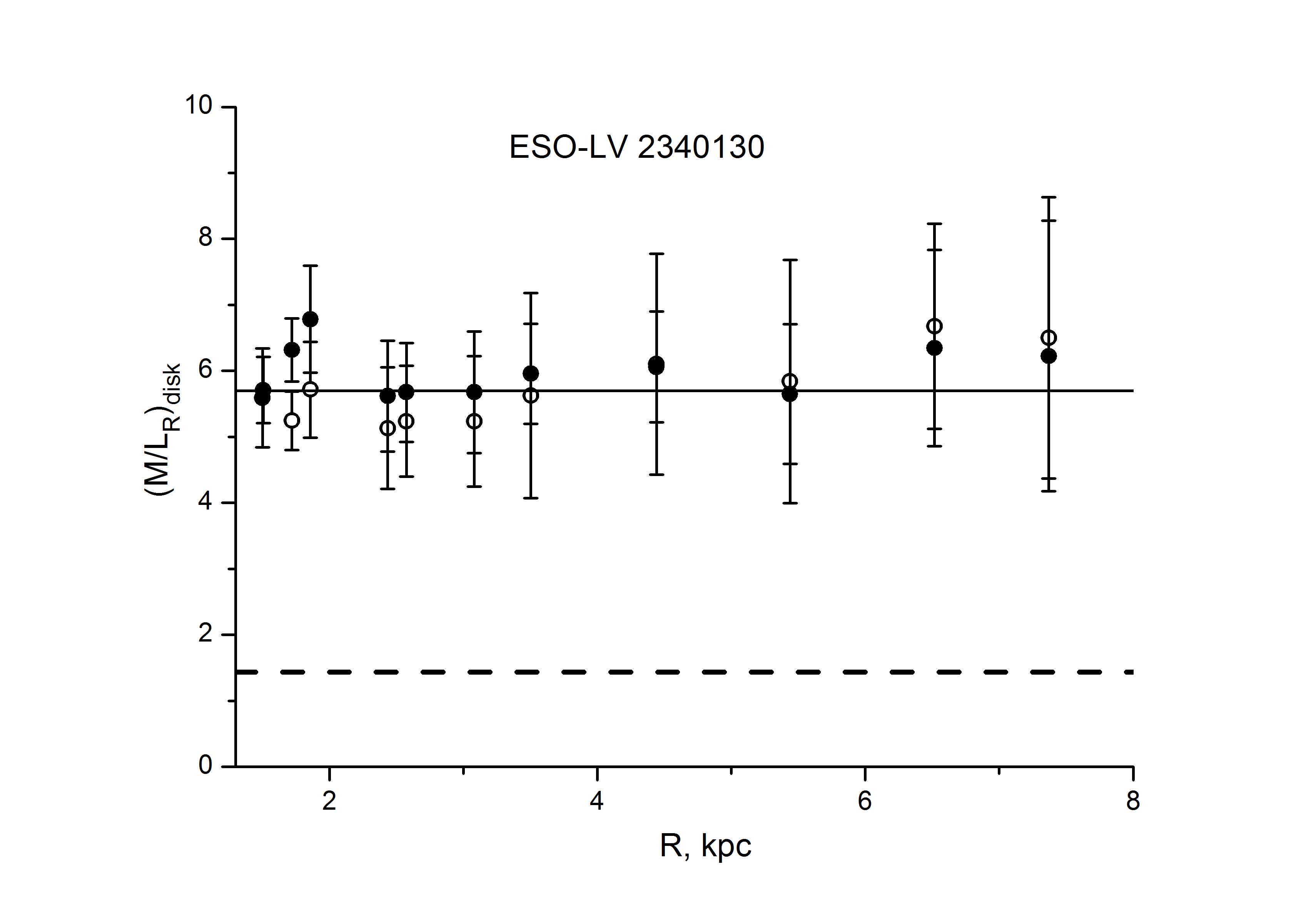}
\includegraphics[width=8cm,keepaspectratio]{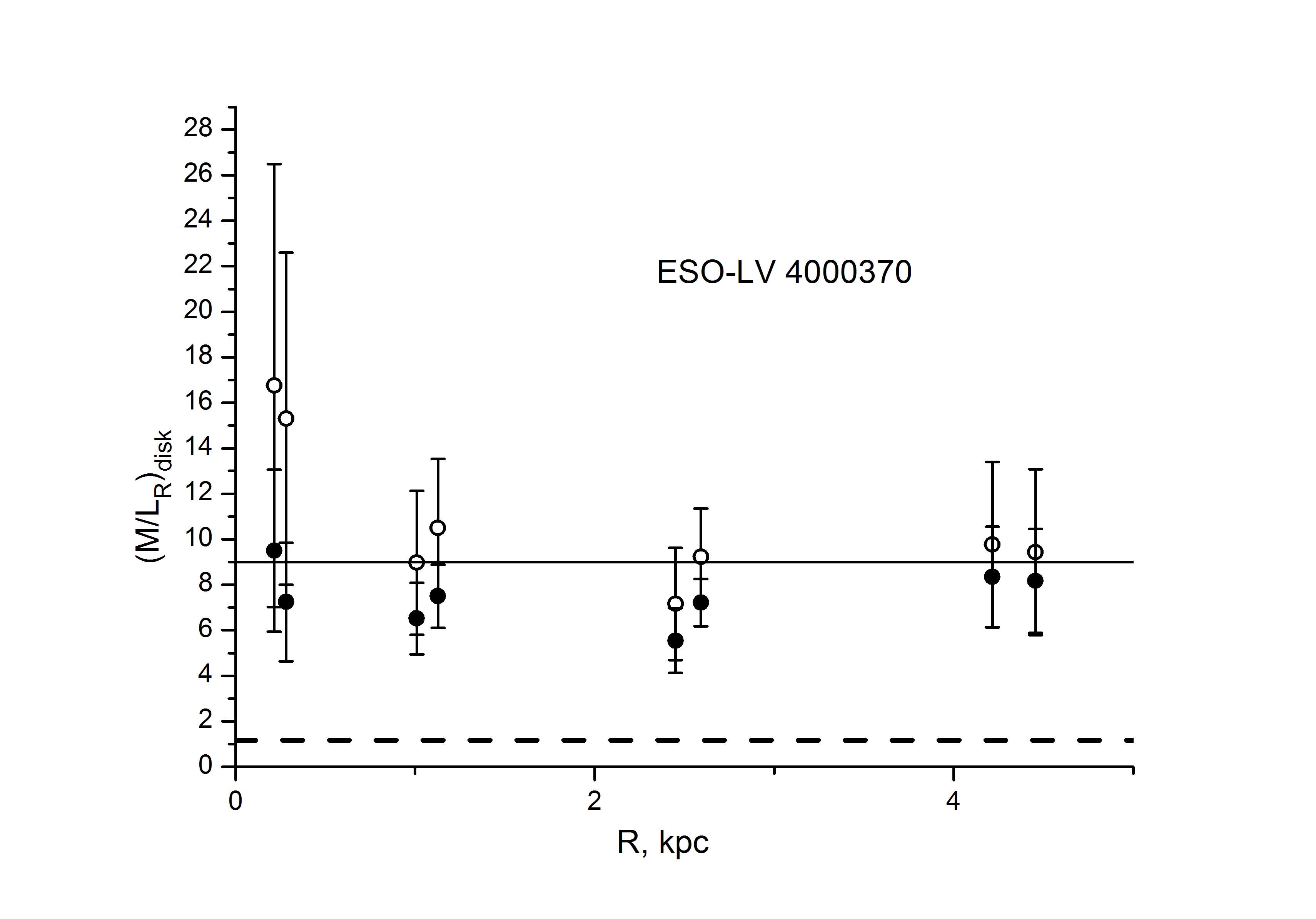}
\caption{Radial profiles of $(M/L_R)_{disk}$ ratios, estimated from the local gravitational stability criterion. Open and filled circles correspond to the gas and stellar circular velocities. The dashed lines denote $(M/L_R)_{disk}$ ratios, predicted by stellar population models (Bell, de Jong, 2001\cite{bdj}). The solid lines correspond to the mean values of $(M/L_R)_{disk}$, which were used for the decomposition of rotation curves.}
\end{figure}

From Fig. 1 it follows that in spite of some differences, the distributions of $(M/L_R)_{disk}$, obtained on the basis of both gas and stellar rotation curves in three of four cases are nearly constant (for ESO-LV2060140 this ratio is roughly constant only in the inner part $R<5$ kpc). Central parts of radial profiles of $(M/L_R)_{disk}$ of the galaxies with significant bulges were excluded from consideration. From Fig. 1 it can be seen, that $(M/L_R)_{disk}$ ratios, corresponding to the stellar disks marginal stability criterion, are several times higher than those predicted by stellar population synthesis models. It is worth reminding, however, that the marginal stability criterion allows one to obtain only the upper limit of disk mass. Nevertheless, the approximate constancy of $(M/L_R)_{disk}(R)$ for marginally stable disks gives evidence that this model may be acceptable for real galaxies.
\section{The rotation curves modeling}
The circular rotation curves of gas and stars were decomposed into the separate components: the exponential disk, the pseudo-isothermal dark halo and, if necessary, King's bulge using the values of $(M/L_R)_{disk}$, found from the gravitational stability condition (see Fig. 1 and Table.\ref{table2}). The radial scalelengths of disks were taken to be equal to the photometrical scalelengths, obtained in R-band by Pizzella et al., 2008 \cite{Pizzella}. The decomposition of the rotation curves of stars after correction for the asymmetric drift (filled circles) and gas (open circles) is illustrated in Fig. 2. For ESO-LV2060140 two modeled rotation curves, corresponding to two different $(M/L_R)_{disk}$ are presented, since one can't unambiguously determine $(M/L_R)_{disk}$ for this galaxy.

\begin{figure}[h!]
\includegraphics[width=8cm,keepaspectratio]{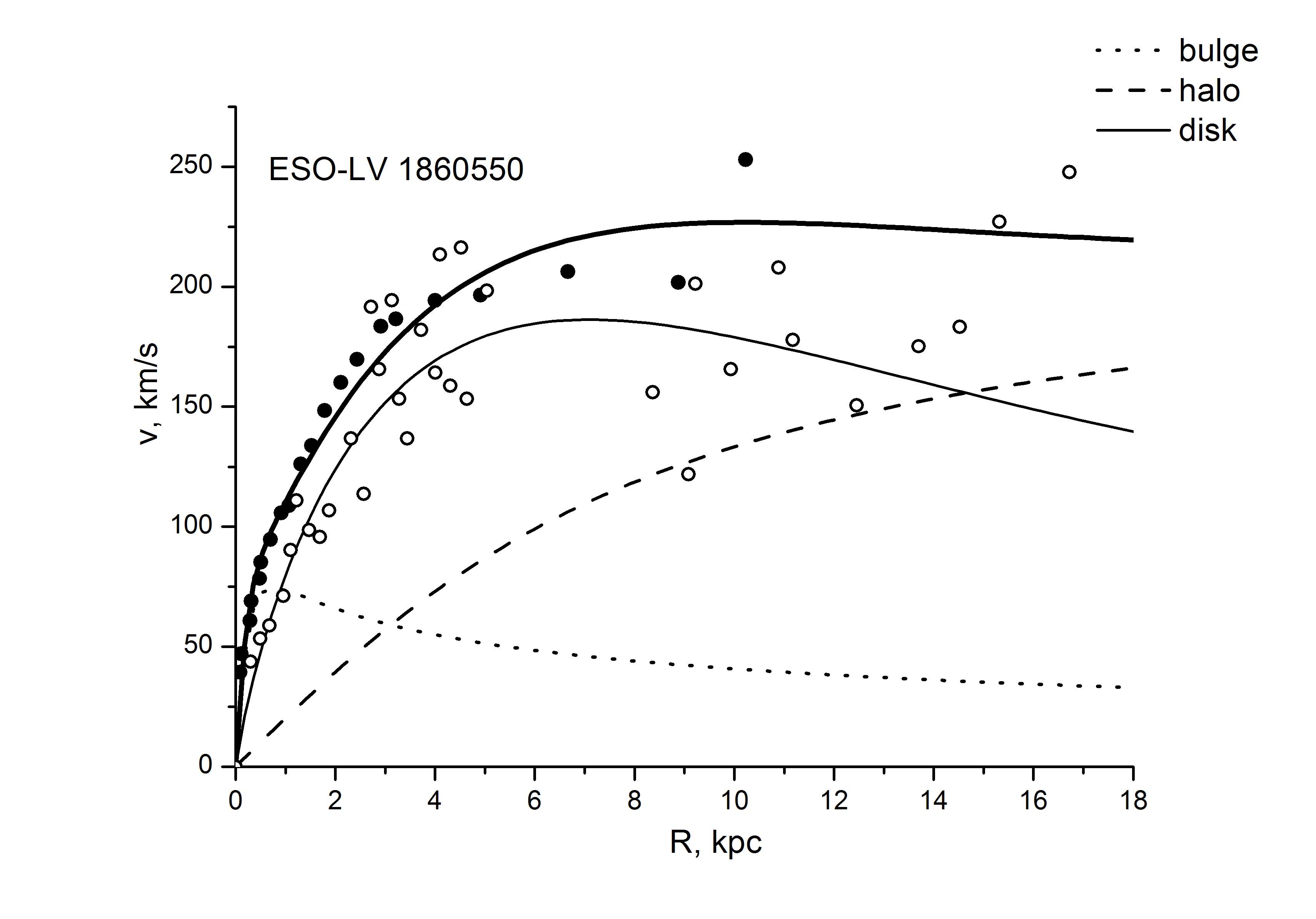}
\includegraphics[width=8cm,keepaspectratio]{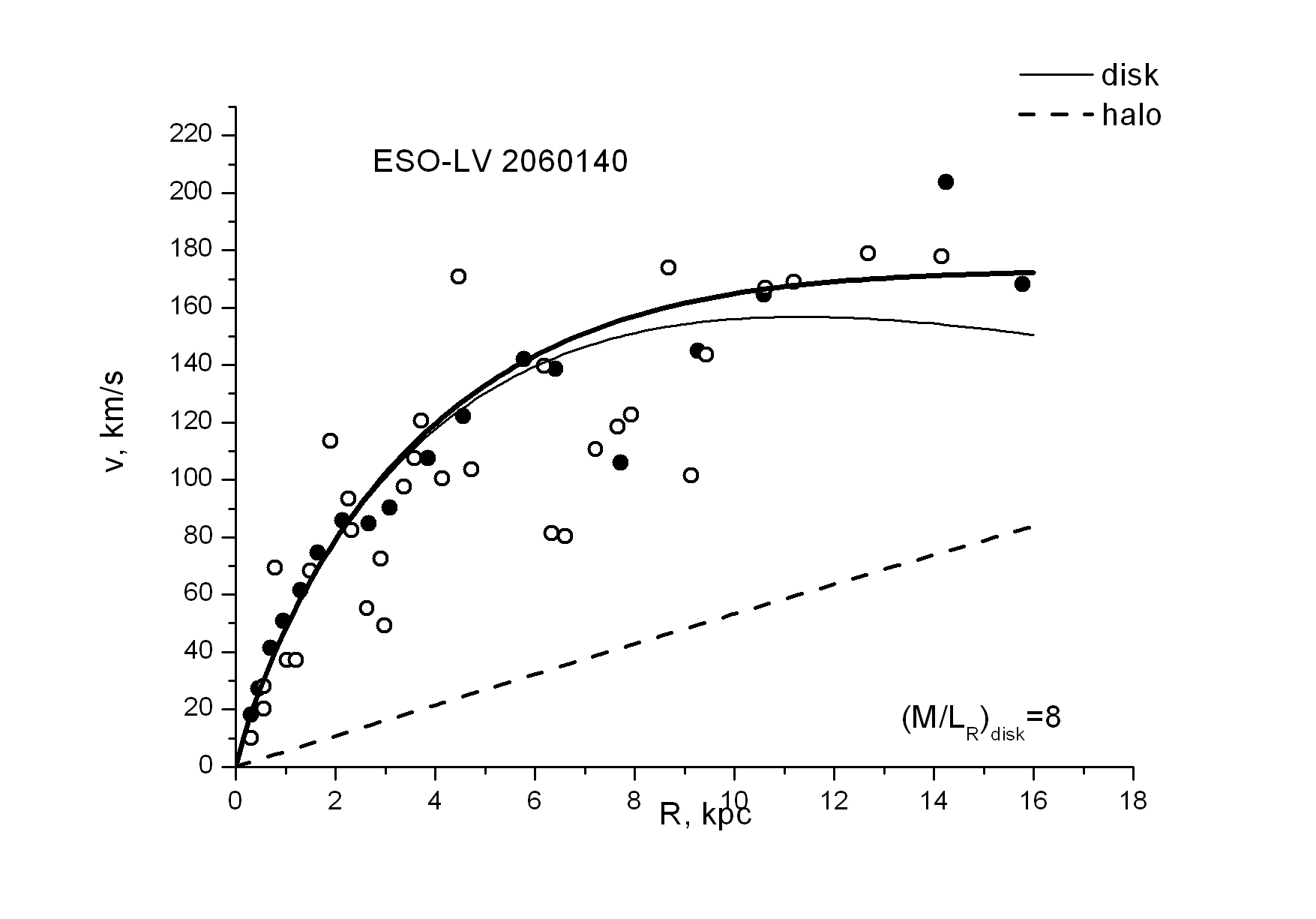}
\includegraphics[width=8cm,keepaspectratio]{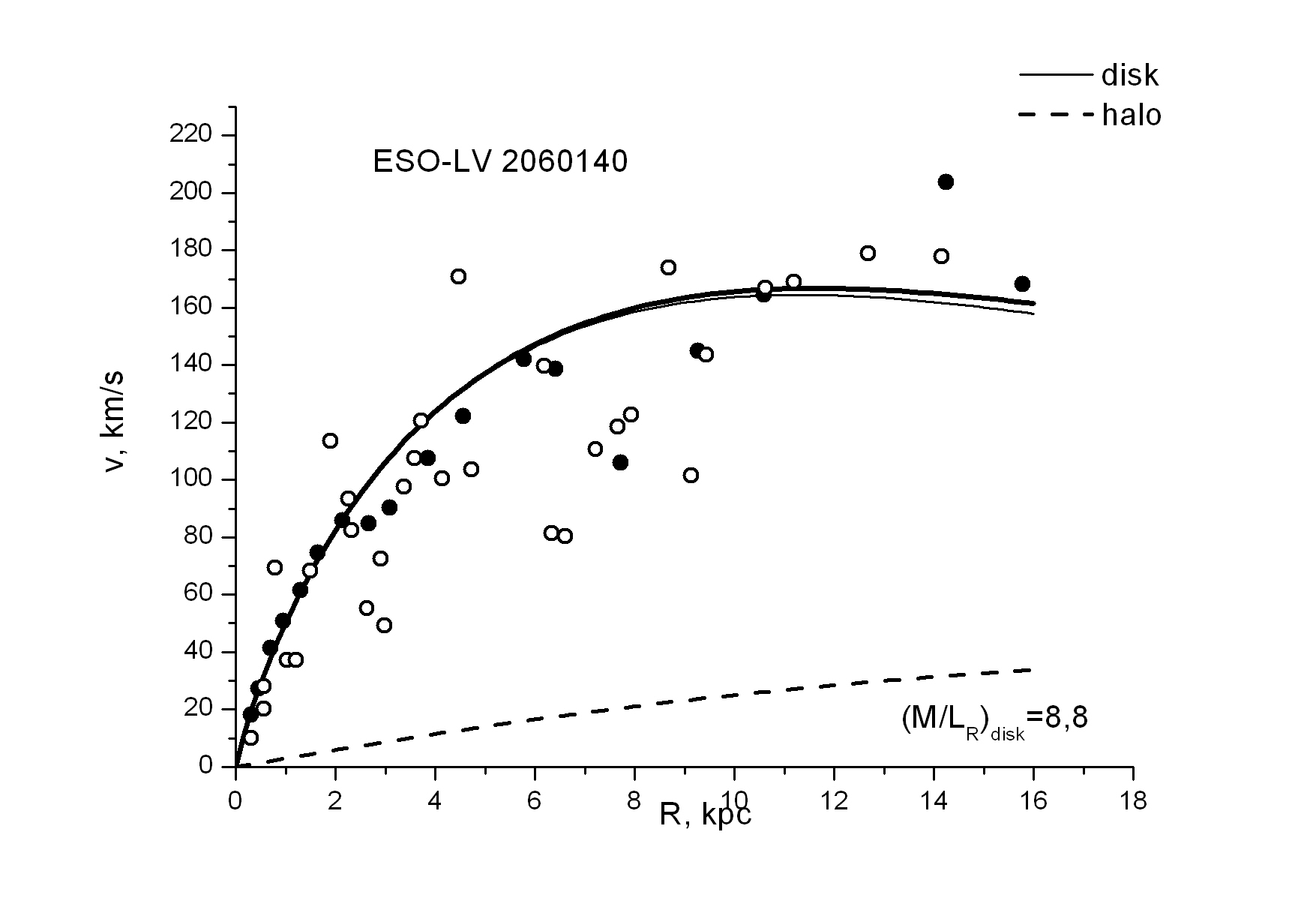}
\includegraphics[width=8cm,keepaspectratio]{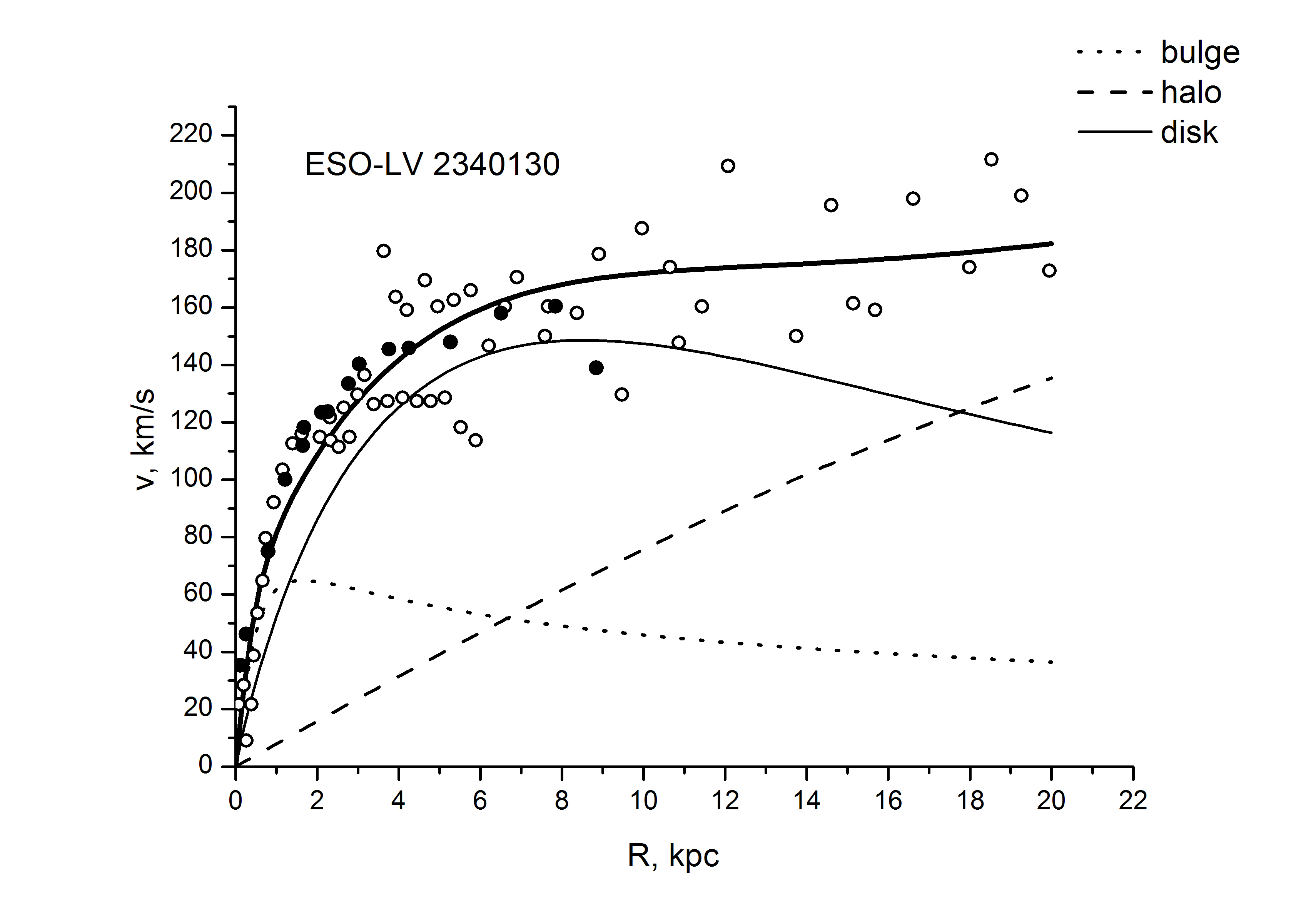}
\includegraphics[width=8cm,keepaspectratio]{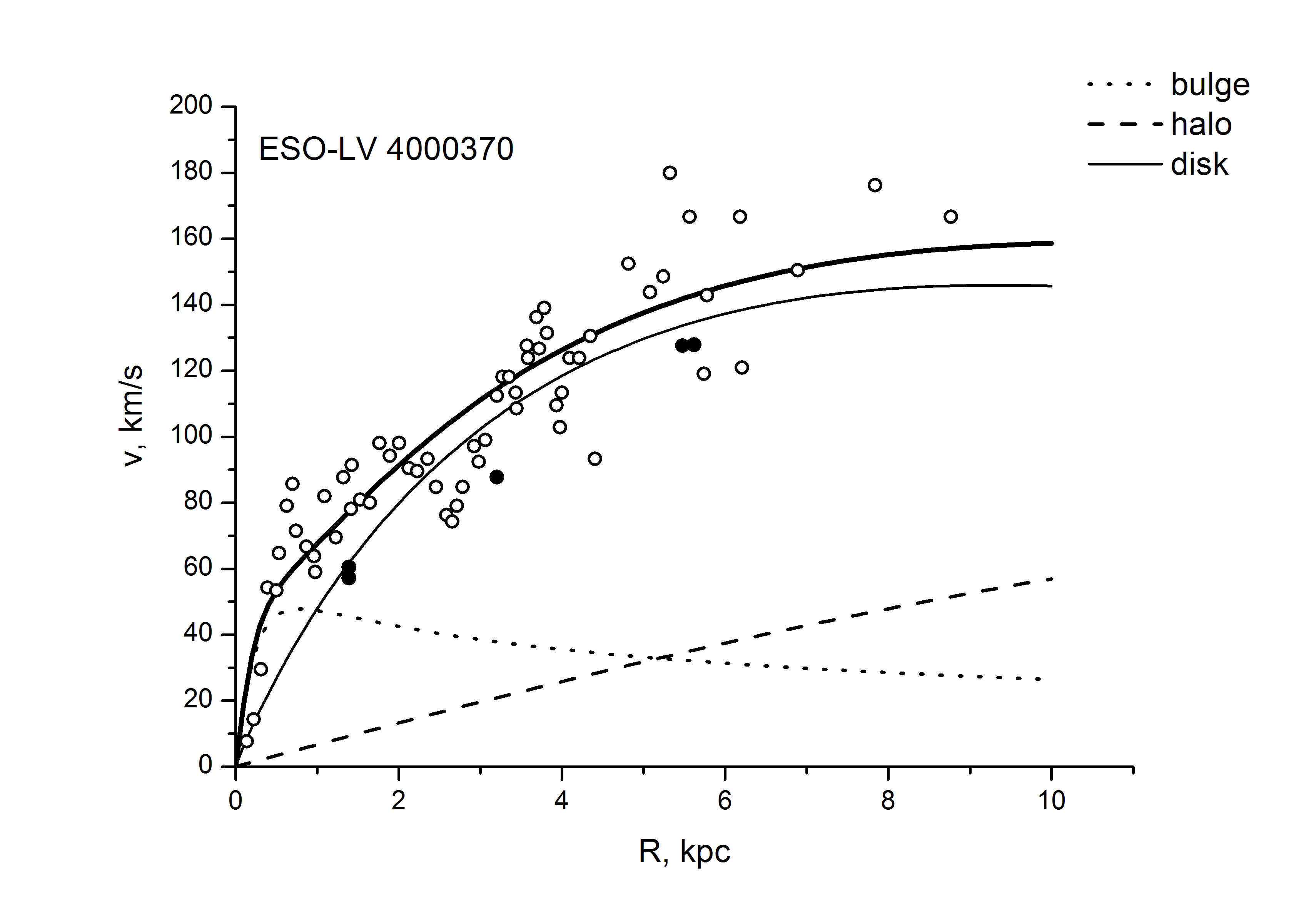}

\caption{The decomposition of the rotation curves of ionized gas (filled circles) and stars (open circles)}
\end{figure}

From Fig. 2 it follows that the models, corresponding to the marginal gravitational stability criterion, provide a reasonable fits to the observed rotation curves.
In Table \ref{table2} the following results of rotation curve modeling are listed (the masses of the components within $R=4h$ were found by means of the extrapolation of the rotation curves to this radius):\\
(1) -- Name\\
(2) -- $(M/L_R)_{disk}$ ratio, which was used for the modeling \\
(3) Ц- The disk mass inside $R=4h$ \\
(4) Ц- The total mass inside $R=4h$ \\
(5) Ц- The dark halo mass inside $R=4h$ \\

\begin{table}[h!]
\small \caption{The results of modeling \label{table2}}
  \begin{center}
    \begin{tabular}{|c|c|c|c|c|c|c|c|c|}
    \hline
Name&$(M/L_R)_{disk}$& $M_{disk} \cdot  10^{10}M_{sun}$ & $M_{tot} \cdot  10^{10}M_{sun}$  & $M_{halo} \cdot  10^{10}M_{sun}$ \\
 (1) & (2) & (3) & (4)&(5) \\
 \hline
ESO-LV 1860550&7,5&6,2&13&6,9\\
\hline
ESO-LV 2060140&8,0&7,0&12&4,9\\
\hline
ESO-LV 2060140&8,8&7,7&8,4&0,7\\
\hline
ESO-LV 2340130&5,7&4,7&9,4&4,1\\
\hline
ESO-LV 4000370&9&5,0&7,5&2,3\\
\hline
 \end{tabular}
  \end{center}
\end{table}

As it follows from Table \ref{table2} the obtained mass of the halo for all considered galaxies is no more than twice as high as that of the disk within $R=4h$. These results indicate that the disks of LSB-galaxies considered here might be significantly more massive, than it can be expected from their low brightnesses for normal stellar population.
 The disk masses of different LSB-galaxies, obtained in this work (filled circles) and by Fuchs, 2003 \cite{Fuchs03} (open circles) are compared with those referred to the maximal disk models in Fig.3. From Fig. 3 it follows that the mass models, obtained in the current work and by Fuchs, 2003 \cite{Fuchs03} are close to the models of maximal disk, where stellar disk has maximal possible contribution to the rotation curve.
\begin{figure} [h!]
\includegraphics[width=12cm,keepaspectratio]{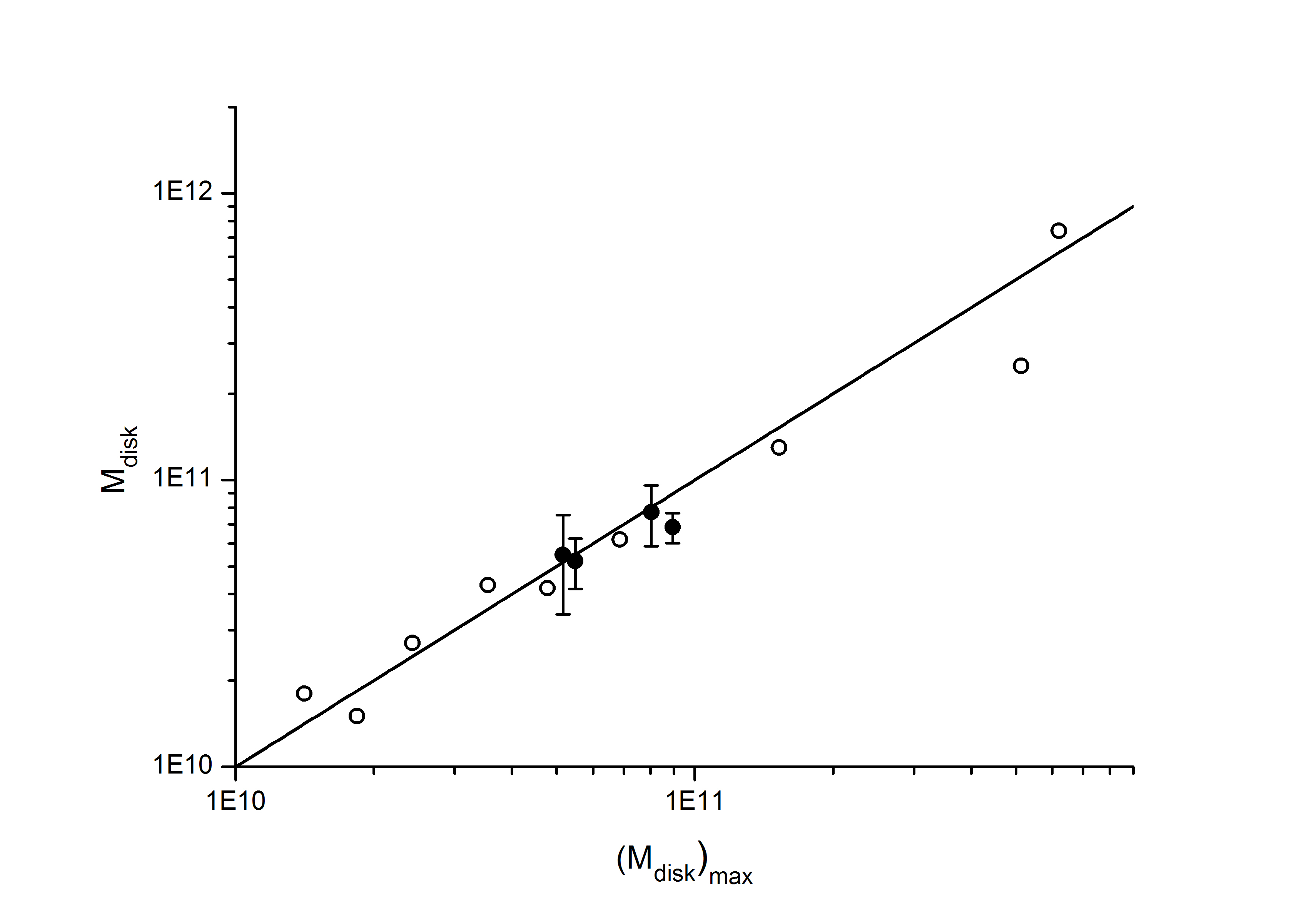}
\caption{Comparison of LSB disk masses, obtained using my models (filled circles) and models of Fuchs, 2003 \cite{Fuchs03} (open circles) with those, referred to the maximal disk models. Black line corresponds to $M_{disk}=(M_{disk})_{max}$.}
\end{figure}

As it was mentioned above, LSB-galaxies are often used to test the models of galaxy formation in the gravitational field of a dark halo by comparing the observed density profiles of these objects with those predicted in terms of cosmological modeling, neglecting the contribution of disk component. The results described above indicate that this approach may not work at least for considered LSB-galaxies. Moreover, according to Kuzio de Naray et al., 2008 \cite{Kuzio08}, the density profile predicted by cosmological modeling (NFW profile) badly enters in the maximal disk models of LSB-galaxies (in other words, the NFW and "max disk" models are incompatible). In this case, if the disks of the galaxies considered here aren't dynamically overheated, and their mass models are really close to the models of maximal disk, it can be an argument against the universality of NFW profile.

\section{On the densities and thicknesses of stellar disks} \label{s5}

The stellar velocity dispersion data allows to obtain not only disk surface densities (or their upper limits), but also half-thicknesses of stellar disks (or their lower limits), corresponding to their stability condition. A half-thicknesses of isothermal self-gravitating, marginally stable disks: $z_{*} = \frac{c_z^2}{\pi G\sigma_*}$ at the fixed distance $R=h$ may be compared with those calculated using the densities, resulted from the photometrical data.
In the later case a self-consistent equilibrium equations were solved for stellar and gaseous components of a disk in the gravitational field of rigid massive spheric pseudo-isothermal dark halo.

A comparison of the disk parameters, obtained using marginal stability condition (column a) with those referred to the stellar population modeling (column b) is presented in Table \ref{table3}:\\
(1) Ц Name\\
(2) -- $z_{*}/h$ ratio at the galactocentric distance $R=h$ \\
(3) Ц- Central surface density of stellar disks $\sigma_{0*}$  \\
(4) Ц The ratio of halo and disk masses inside $R=4h$  \\
(5) Ц- Radial scalelenghts of the disks\\

\begin{table}[h!]
\small \caption{ A comparison of parameters estimated from gravitation stability criterion and photometric models. \label{table3}}
    \begin{tabular}{|p{3cm}|p{1,5cm}|p{1,5cm}|p{1,5cm}|p{1,5cm}|p{1,5cm}|p{1,5cm}|p{1,5cm}|p{0,4cm}|}
    \hline
Name& \multicolumn{2}{c|}{$z_{*}/h$, pc}& \multicolumn{2}{c|}{$\sigma _{0*}$, $M_{sun}/$pc$^2$}& \multicolumn{2}{c|}{$M_{halo}/M_{disk}$}& \multicolumn{1}{c|}{$h$, kpc}  \\
(1) & \multicolumn{2}{|c|}{(2)} & \multicolumn{2}{c|}{(3)} & \multicolumn{2}{c|}{(4)}&(5) \\
 \hline
&a &b& a & b& a &b & \\
\hline
ESO-LV 1860550&0,1&0,2&1000&240&1,1&11&3,6\\
\hline
ESO-LV 2060140&0,2&0,4&450&68&0,7&9,1&5,3\\
\hline
ESO-LV 2340130&0,1&0,3&590&150&0,9&7,4&3,8\\
\hline
ESO-LV 4000370&0,2&0,4&520&85&0,5&26&4,1\\
\hline
 \end{tabular}
\end{table}

For ESO-LV2060140 in case (a) the value $(M/L_R)_{disk}=8$ was used.
From Table \ref{table3} it can be seen, that the photometric models lead to the higher contribution of dark halo to the rotation curve in comparison with marginally stable model. The disk surface densities, estimated for normal stellar population models give 2-2,5 times thicker disks in comparison with the marginal stable ones. The estimates of $z_{*}/h$, corresponding to the marginal stability criterion, better agree with those expected for LSB galaxies from the relationship between the deprojected surface brightness and disk thickness, found by Bizyaev, Mitronova, 2009 \cite{Bizyaev}, where the objects with lower surface brightness have lower $z_{*}/h$.

\section{Conclusions}

The disk masses of four LSB-galaxies, obtained using the local gravitational stability criterion, appear to be significantly higher, than it can be expected from their photometrical data in the assumption of normal stellar population. However they agree with the observed rotation curves. The corresponding mass models are close to the models of maximal disk.
Hence, either the disks of these galaxies are strongly dynamically overheated and thick, or their densities and masses are quite typical for HSB-galaxies. In this case their high mass-to-light ratios can indicate that either they contain a large amount of low massive stars (a heavy bottom IMF), or a significant fraction of dark matter is contained in the disks, either in baryonic (e.g. cold molecular gas, see Pfenniger, Combes, 1994 \cite{pfenniger}) or non-baryonic form.

It seems that at least some LSB-galaxies are not dark halo dominated systems which possess the disks of low density. In this case LSB galaxies differ from HSB ones not by the high contribution of dark halo, but rather by peculiarities of their disks in the presence of dark halo of moderate mass.

\section*{Acknowledgments}
I would like to thank A.V. Zasov for the initiating this work and for helpful discussions. I also want to thank O.V. Abramova for carrying out the calculation of the thicknesses of stellar disks in the gravitation potential of dark halo.

I acknowledge the possibility to use the HyperLeda database (http://leda.univ-lyon1.fr/).

This work was partly supported by Russian Foundation for Basic Research, grant 07-02-00792.

\end{document}